# Large single crystal growth of BaFe$_{1.87}$Co$_{0.13}$As$_2$ using a nucleation pole

D. L. Sun, Y. Liu, J. T. Park and C. T. Lin*

*Max-Planck-Institut für Festkörperforschung, Heisenbergstr. 1, D-70569 Stuttgart, Germany*

(Date: August 10 2009)

Co-doped iron arsenic single crystal of BaFe$_{1.87}$Co$_{0.13}$As$_2$ with dimension up to 20 × 10 × 2 mm$^3$ was grown using a nucleation pole: an alumina stick served as nucleus center during growth. The high quality of crystalline was illustrated by the measurements of neutron rocking curve and X-ray diffraction pattern. A very sharp superconducting transition temperature $T_c$~25 K was revealed by both resistivity and susceptibility measurements. A nearly 100% shielding fraction and bulk nature of the superconductivity for the single crystal were confirmed using magnetic susceptibility data.

## I. INTRODUCTION

The newly discovered iron-arsenic superconductors [1] have stimulated a great interest in the study of their structural, transport and magnetic properties. The parent compounds show a spin-density-wave (SDW) ground state. With introducing charge carriers by electron or hole doping, SDW transition is suppressed while superconducting transition takes place, being reminiscent of high $T_c$ cuprate superconductors. For instance, the compounds of RO$_{1-x}$F$_x$FeAs (R=La, Pr, Sm, Ce, Nd, and Gd) have shown their superconducting transition with the fluorine substitution for oxygen, resulted in the highest $T_c$ at 56 K [1-7] by electron doping. In case of hole doping, La$_{1-x}$Sr$_x$OFeAs shows relative lower $T_c$. For the oxygen-free compounds, $T_c$ reaches 38 K for hole-doped A$_{1-x}$K$_x$Fe$_2$As$_2$ (A=Ba, Sr, Ca) and 22K for electron-doped AFe$_{2-x}$M$_x$As$_2$ (M=Co or Ni) [8-15]. The superconductivity is suggested to compete with SDW order and spin fluctuations could play a role in forming the Cooper pairs. In addition, the compounds display anisotropic behavior owing to the layered structure when measuring upper critical field [16], coherence length and penetration depth [17] in the plane and along the *c* axis. Therefore, the investigations of spin dynamics are highly attractive to explore the superconducting mechanism of iron-arsenic superconductors. High quality and large single crystals are desperately required for various experiments, particular in the neutron scattering measurements.

Flux method could be used to grow single crystals of iron-arsenic superconductors [2, 10-15]. Several groups reported the growth of AFe$_2$As$_2$ single crystals using KCl/NaCl, Sn or self-flux technique, obtaining the size as large as 12 × 8 × 1 mm [18]. However, the crystals are contaminated by Sn about 1-2 at. % [10]. It has been proved that the Sn is incorporated into the site of Fe detected by the atomic spectroscopy and leads to a decrease of the SDW transition temperature [11]. Therefore, the self-flux method is preferred for the growth to prevent the crystal from contaminations [12-15]. In this work, we present the growth of $BaFe_{2-x}Co_xAs_2$ single crystals using an alumina stick as nucleation center in melt during growth. Large crystal disk sized up to $\phi 40 \times 5$ mm were obtained. The structure characterizations with neutron rocking curve and X-ray diffraction illustrate high quality of the samples. The measurements of resistivity and magnetic susceptibility show a sharp superconducting transition temperature at ~25 K.

---

* Corresponding author; E-mail address: ct.lin@fkf.mpg.de

## II. EXPERIMENT

The starting materials in the formula of $BaFe_{5-x}Co_xAs_5$ (all elements are from Alfa Aesar, 4-5 N in purity) with x=0.5 were used for the self flux growth. Usually ~20 g of the mixtures were well ground and then loaded in a $ZrO_2$ crucible covered with a lid to minimize the arsenic volatilization. An $Al_2O_3$ stick of $\phi 2 \times 70$ mm was inserted in a hole drilled through the lid and dips into the mixtures to serve as a nucleation pole. All preparation procedures were carried out in a glove box containing Ar. The loaded crucible was then sealed in a quartz ampoule filled with 250 mbar argon atmosphere. The ampoule was placed in a furnace and heated up to 1190 ℃ for 10 h. The temperature of the melt was then lowered to 1090 ℃ at a rate of ~2 ℃/h, followed by decanting the residual flux. Finally the furnace was cooled down to room temperature at 100 ℃/h. The whole procedure of crystal growth was carried out in a sealed system with a specially designed apparatus, as shown in Fig. 1.

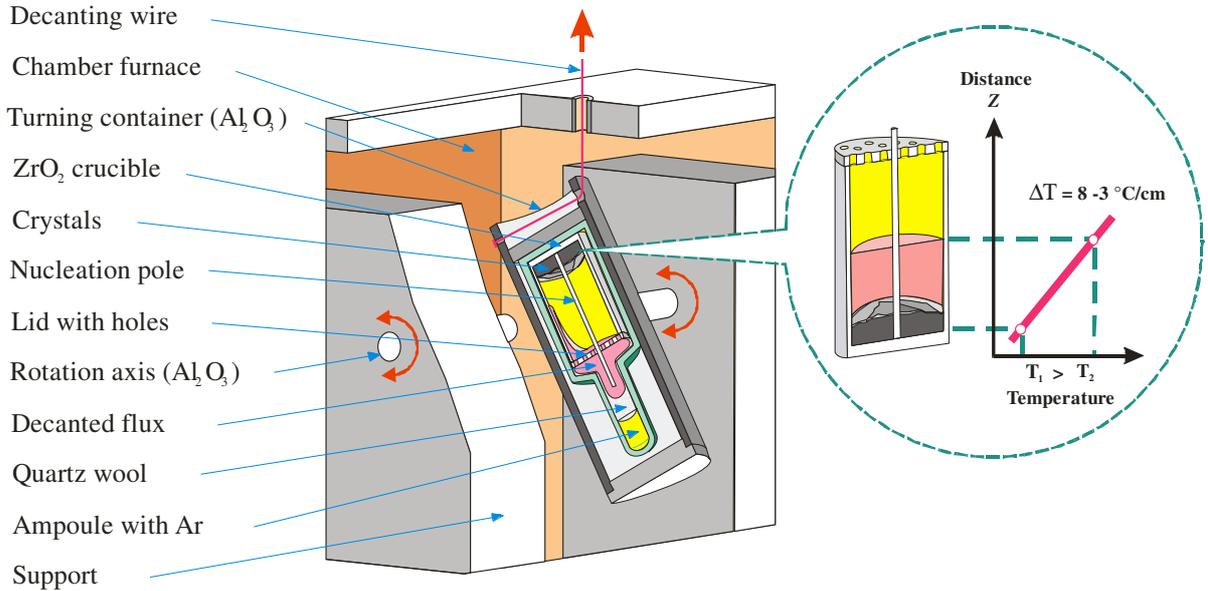

Fig. 1 A schematic drawing of the apparatus used to grow single crystals of $BaFe_{1.87}Co_{0.13}As_2$. The inset shows a temperature gradient of 8-3 °C/cm distributed from the bottom to the upper part of solution.

X-ray diffraction (XRD) measurements of single crystals were carried out with the X-ray diffractometer (Philips PW 3710) using Cu Kα radiation and with a scanning rate of 0.02 °/min. The lattice parameters derived from the powder XRD using TOPAS v2.1, which is a general profile and structure analysis software for powder diffraction data (Karlsruhe, Germany,

2002). Energy dispersive X-ray spectroscopy (EDX) analyses were employed to determine the crystal composition. The error of the analysis is within 1 at%. Neutron data were collected using the thermal triple-axis spectrometer PUMA at the research reactor FRM-II, Garching, Germany. The initial neutron wavelength was fixed to 2.66 Å and a flat PG (002) monochromator was used. The resistivity was obtained using a physical property measurement system (PPMS$^{TM}$, Quantum Design). DC susceptibility was obtained by a SQUID (VSM, Quantum Design) magnetometer.

### III. RESULTS AND DISCUSSION

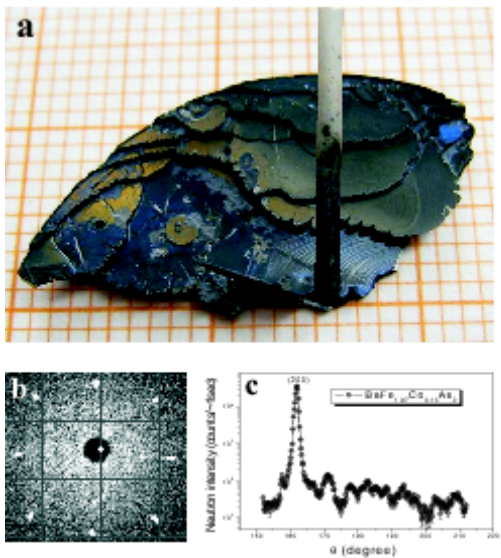

Fig. 2 (a) A part of BaFe$_{1.87}$Co$_{0.13}$As$_2$ crystals with mass of ~2 g and the alumina "seed" used to locate in the centre of the crystal disk. (b) The related Laue pattern showing the (00l) face and (c) Neutron rocking curve of the crystals (a), the weaker diffraction peaks obtained from the mis-orientated small crystals which adhered on side of the big one during decanting.

A nucleation center is crucial for the growth of a large and high quality single crystal. In our growth experiment, we used an alumina stick served as a "seed" to play a role of nucleation pole in the melt during growth. The alumina stick was positioned at the bottom of the crucible. This allows the heat convection transports from hotter melt $T_1$ to colder end $T_2$ of the stick, creating a temperature gradient of ~8-3 °C/cm when the heating temperature ranges between 1190 and 1090 °C. The schematic drawing of the experiment set up is shown in Fig. 1. By the use of this "seeding" method together with a low cooling rate of ~2 °C/h applied, spontaneous and numerous nuclei can be minimized during growth. The process of crystallization takes place around the "seed" of which is the colder pole of the stick. With cooling the crystals gradually grow up, as result in a large crystal together with some small crystals formed around the big one. We demonstrate that the biggest crystal can be grown as large as $\phi 40 \times 5$ mm, which thickness depends on the amount of source materials. Fig. 2(a) shows the "seed" rod locating in the center of the crystal disk and the broken part of the crystal measured up $20 \times 10 \times 2$ mm$^3$, whose size is one third of the disk. This is the largest crystal of iron-arsenic superconductor ever reported [2, 10-15]. It is noticed that to obtain crystal free from flux a temperature of ~1090 °C had to be maintained for ~2 h after decanting and then cooling to room temperature. This allows the residual flux flowing out completely to leave free-standing crystals inside of the crucible. It is emphasized that the decanting device is specially designed with a movable nickel wire for tilting the crucible on its top and drop the residual flux at high decanting temperatures out of the furnace. This can avoid any poisoning from arsenic in case of a crack of the quartz tube occurred.

The crystals exhibit layered structure of the $c$ plane. The space between layers evidences the room filled with liquid flux during growth and the crystals are free standing after decanting. All layers are the (001) planes and belong to the tetragonal structure characterized by the X-ray Laue pattern in Fig. 2(b). The composition of the crystal was determined to be $BaFe_{1.87}Co_{0.13}As_2$, which content of the cobalt is approximately one quarter of the initial doping amounts. According to six doping levels of Co (x=0.25, 0.35, 0.50, 0.55, 0.70 and 0.80 in starting materials; x= 0.08, 0.11, 0.13, 0.18, 0.20 and 0.24 in crystals) in our experiment, the segregation coefficient K is estimated to be ~0.71. It is indicative of, K<1, the solubility of Co in the crystal is low and therefore a higher doping content in solution results in a lower in crystals. The in-plane and out-of-plane crystallization of the whole bulk $BaFe_{1.87}Co_{0.13}As_2$ crystals were revealed by the neutron rocking curve, as shown in Fig. 2(c). The (200) reflection is pronounced and symmetrical. The FWHM is around 1.01º (in-plane) and 1.083º (out-of-plane). Both peaks are very sharp and performed within the instrumental resolution limited. Therefore the real peak width (mosaic of the samples) could be even narrower. It is indicative of an excellent quality for the whole crystal.

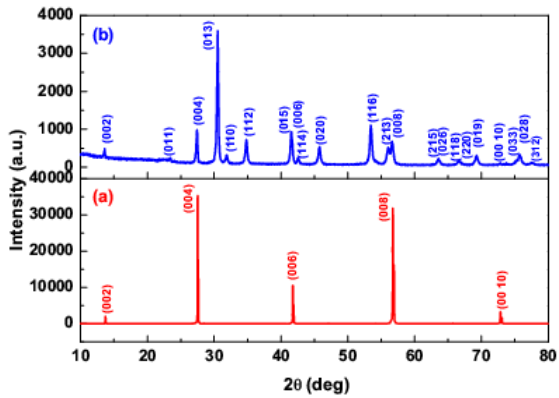

Fig. 3 X-ray diffraction patterns of $BaFe_{1.87}Co_{0.13}As_2$ (a) the cleaved plane of a crystal and (b) the powders of a ground crystal.

The structure of the $BaFe_{1.87}Co_{0.13}As_2$ single crystal was identified by XRD measurements. The XRD diffraction patterns were obtained using ground powders of as-grown crystals. Fig. 3(a) shows the *(00l)* reflections obtained from the naturally cleaved surface, indicating a good crystallization along the *c* axis of the crystal. The powder XRD pattern can be well indexed by the space group I4/mmm, as shown in Fig. 3(b). The lattice parameters $a$=3.9633(8) Å and $c$=12.991(2) Å, derived from the program using TOPAS 2.1. Compared to the undoped $BaFe_2As_2$ with $a$=3.9635(5) Å, $c$=13.022(2) Å [14], the *c* lattice parameter of the $BaFe_{1.87}Co_{0.13}As_2$ slightly shrinks for 0.24 % while the *a* is almost unchanged. This is owing to the ionic radius of $R_{Co}^{3+}$(0.61 Å)< $R_{Fe}^{2+}$ (0.78 Å), since the $Co^{3+}$ ions partially substitutes for $Fe^{2+}$ in the basal plane of $BaFe_{1.87}Co_{0.13}As_2$. Our results are consistent with the reported data [14].

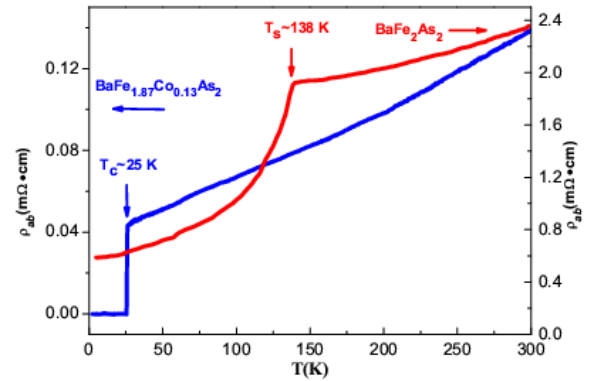

Fig. 4 Temperature dependence of resistivity for single crystals of $BaFe_2As_2$ obtained by self-flux method [11] and $BaFe_{1.87}Co_{0.13}As_2$.

The temperature dependence of resistivity for $BaFe_2As_2$ and $BaFe_{1.87}Co_{0.13}As_2$ single crystals in the temperature range of 2-300 K is shown in Fig. 4. For the pure $BaFe_2As_2$, the resistivity shows a rapid decrease below $T_s$~138 K, corresponding to SDW transition. In case of $BaFe_{1.87}Co_{0.13}As_2$ crystal, a very sharp superconducting transition occurs at $T_c$~25 K, while the resistivity kink of SDW disappears in the Co-doped sample.

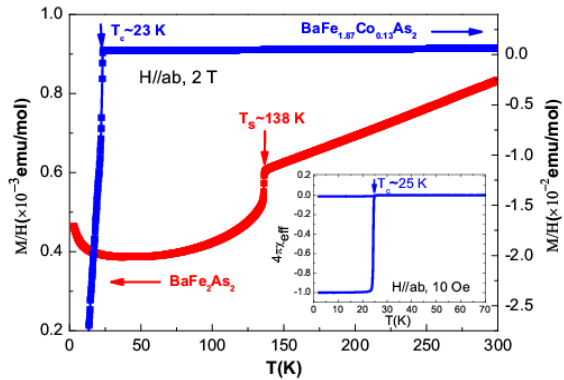

Fig. 5 Temperature dependence of the magnetic susceptibility of $BaFe_2As_2$ [11] and $BaFe_{1.87}Co_{0.13}As_2$, respectively. Inset is the measurement performed on both zero field cooling (ZFC) and field cooling (FC) modes under 10 Oe.

Fig. 5 shows the temperature dependence of magnetic susceptibility of $BaFe_2As_2$ and $BaFe_{1.87}Co_{0.13}As_2$ crystals at 2 T field and the temperature range of 2-300K. A kink associated with SDW transition is observed in $BaFe_2As_2$ crystal [10, 11], consistent with the result of resistivity measurement. For $BaFe_{1.87}Co_{0.13}As_2$, the SDW transition is entirely suppressed with the occurrence of superconducting transition. There is no sign of SDW transition with an application of 2 T magnetic fields. A sharp transition occurs at $T_c \sim 25$ K for the $BaFe_{1.87}Co_{0.13}As_2$ crystal, as shown in the inset of Fig. 5. The shielding fraction close to 1 demonstrates a bulk nature of the superconductivity. The onset transition temperature $T_c \sim 25$ K is defined as magnetic susceptibility in normal state decreases 10%, which is the highest one ever reported. The superconducting transition width is 0.5 K with $\Delta T_c = T_c(10\%) - T_c(90\%)$, which is close to the reported 0.6 K ($T_c=22$ K) for $BaFe_{1.8}Co_{0.2}As_2$ [14], and much smaller than 2.3 K ($T_c=14.3$ K) for $LaFe_{0.92}Co_{0.08}AsO$ [19], 2.5 K ($T_c=36.3$ K) for $Ba_{0.6}K_{0.4}Fe_2As_2$ [12] and 4.5 K ($T_c=28.2$ K) for $LaFeAsO_{0.89}F_{0.11}$ [20], respectively.

## IV. CONCLUSIONS

In summary, the large and high quality $BaFe_{1.87}Co_{0.13}As_2$ single crystals have been grown successfully by the "seeding" method. The neutron rocking curve and X-ray diffraction patterns show that the good crystallization of the sample. The resistivity and susceptibility results show sharp superconducting transitions at 25 K and nearly 100% shielding fraction, which confirms the bulk nature of the superconductivity and high quality of single crystal.

## ACKNOWLEDGMENTS

We thank G. Götz for the XRD measurements. P. Popovich for the SQUID performance, C. Busch for the analyses of crystal composition, and L. Dorner-Finkbeiner and H. Bender for the technique support. K. Hradil at FRM-II for neutron rocking curve performance.